\documentclass{elsart}

\usepackage{graphicx,amssymb,times,amsmath,amsfonts,epsfig}

\def\be{\begin{eqnarray}}
\def\ee{\end{eqnarray}}

\def\b{\bibitem}

\begin{document}
\begin{frontmatter}
\title{A theorem for the existence of Majorana fermion modes in spin-orbit-coupled semiconductors}

\author[Clemson]{Sumanta Tewari}
\ead{stewari@clemson.edu}
\author[CMTC]{Jay D. Sau}
\author[CMTC]{S. Das Sarma}

\address[Clemson]{Department of Physics and Astronomy, Clemson University, Clemson, SC 29634}
\address[CMTC]{Condensed Matter Theory Center and Joint Quantum Institute, Department of Physics, University of
Maryland, College Park, Maryland 20742-4111}

\begin{abstract}
We prove an index theorem for the existence of Majorana zero modes in a semiconducting thin film with a sizable spin-orbit coupling when it is adjacent to an $s$-wave superconductor. The theorem, which is analogous to the Jackiw-Rebbi index theorem for the zero modes in mass domain walls in one-dimensional Dirac theory, applies to vortices with odd flux quantum in a semiconducting film in which $s$-wave superconductivity and a Zeeman splitting are induced by proximity effect. The momentum-space construction of the zero-mode solution presented here is complementary to the approximate real-space solution of the Bogoliubov-de Gennes equations at a vortex core [J. D. Sau \emph{et al.}, arXiv:0907.2239], proving the existence of non-degenerate zero-energy Majorana excitations and the resultant non-Abelian topological order in the semiconductor heterostructure. With increasing magnitude of the proximity-induced pairing potential, the non-Abelian superconducting state makes a topological quantum phase transition to an ordinary $s$-wave superconducting state which no topological order.
\end{abstract}

\end{frontmatter}

\section{Introduction}

Particle statistics is a genuinely quantum mechanical concept which has no classical analog. In spatial dimension three and above,
the wave function of a many-body quantum state of identical particles remains either unchanged (bosons) or undergoes a change of sign (fermions) under a pairwise interchange of the particle coordinates. That there are only two possibilities is a consequence of the fact that there are only two irreducible representations of the permutation group for $N$ particles. However, in $(2+1)$ dimension, where permutation and exchange are not necessarily equivalent, the quantum statistics of particles can be remarkably different from the ordinary statistics of bosons and fermions \cite{Leinaas, Wilczek}. In this case, under a pairwise interchange of the particle coordinates, it is possible for the many body wave function to receive an arbitrary phase factor $e^{i\theta}$ where $\theta$ is an angle intermediate between $0$ (boson) and $\pi$ (fermion). The particles obeying statistics given by the angle $\theta$ are called anyons \cite{Wilczek2}. Even if the anyon statistics is remarkably different from that of bosons and fermions, since the simple phase factor is only a one-dimensional representation of the braid group in 2D, the statistics is still Abelian. A more exotic possibility, one which has remarkable prospect for fault-tolerant topological quantum computation (TQC) \cite{Nayak-RMP}, arises when the quantum ground state of the many-particle system is degenerate. In situations where the many body ground state wave function is a linear combination of states from this degenerate ground state subspace,
a pairwise exchange of the particle coordinates can unitarily \emph{rotate} the wave-function in the subspace.
In this case, the exchange statistics is given by a multi-dimensional unitary matrix representation
 of the 2D braid group, and, thus, the statistics is non-Abelian.
It has been proposed \cite{Kitaev} that such a system, where the ground state degeneracy is protected by a gap from
local perturbations, can be used as a fault-tolerant platform for TQC.

Recently, some exotic ordered states in condensed matter systems, such as the Pfaffian states in
fractional quantum Hall (FQH) systems~\cite{Moore, Nayak-Wilczek, Read, dassarma_prl'05} and
chiral $p$-wave superconductors/superfluids~\cite{Ivanov, Stern, DasSarma_PRB'06, tewari_prl'2007},
as well as the surface state of a topological insulator (TI) in which $s$-wave superconductivity is induced by proximity effect~\cite{fu_prl'08, fu_prl'09, akhmerov_prl'09}, have been proposed
as systems which support quasiparticles with non-Abelian statistics, and, therefore, can potentially be used as TQC platforms. The common thread between these systems is that they all allow quasiparticle excitations which involve no energy cost (when the mutual separation among the excitations is large). The second quantized operators, $\gamma_i$, corresponding to these zero energy excitations are self-hermitian, $\gamma_i^{\dagger}=\gamma_i$, which is in striking contrast to ordinary fermionic (or bosonic) operators for which $c_i \neq c_i^{\dagger}$. However, since $\gamma_i$ and $\gamma_j$, which are called Majorana fermion operators, anticommute when $i \neq j$, they retain some properties of ordinary fermion operators as well. It is the self-hermitian property of the Majorana operators which lies at the heart of the ground state degeneracy and the resulting non-Abelian statistics \cite{Nayak-Wilczek, Ivanov} of quasiparticle excitations in these systems.

It has been shown recently \cite{Sau} that even a regular semiconducting film with a sizable spin-orbit coupling,
such as InGaAs thin films, can host, under suitable conditions, Majorana fermion excitations localized near defects. By an analysis of the real-space Bogoliubov-de Gennes (BdG) equations for a vortex in the semiconductor, in which $s$-wave superconductivity and a Zeeman splitting are proximity induced (Fig. (1a)), it has been shown that the lowest energy quasiparticle excitation is a zero-energy Majorana fermion mode. The Majorana mode is separated by a finite energy gap (so-called mini-gap) from the other conventional fermionic excited states in the vortex core. Thus, for a collection of well-separated vortices, the resulting degenerate ground state subspace is protected from the environment by the mini-gap, enabling the potential use of the semiconductor heterostructure in Fig. (1a) in TQC. Since the basic effects behind the emergence of the Majorana fermion excitations -- spin-orbit coupling, $s$-wave superconductivity, and Zeeman splitting -- are physically well-understood and experimentally known to occur in many solid state materials, the proposed semiconductor heterostructure \cite{Sau} is possibly one of the simplest systems to realize, which supports non-Abelian topological order.

In this paper, we show that the existence of the Majorana fermion zero modes at the cores of the vortices in the semiconductor heterostructure in Fig. (1a) is due to an index theorem. The theorem is analogous to the one proven by Jackiw and Rebbi \cite{Jackiw1, Jackiw2} for the existence of zero energy eigenstates at mass domain walls in a 1D system of Dirac fermions, which was later applied \cite{Sumanta} to prove the existence of zero-energy excitations at vortices in a spinless 2D chiral $p$-wave superconductor. In analogy with the chiral $p$-wave superconductor, we find that while an odd flux-quantum vortex in the semiconductor thin film traps a unique zero energy eigenstate, an even flux-quantum vortex does not. Furthermore, since the components of the fermion field in our analog of the 1D Dirac theory are pairwise related by parity reversal and hermitian conjugation, we explicitly show that the second quantized operator corresponding to the zero-energy eigensolution is self-hermitian, $\gamma^{\dagger}=\gamma$. This is in contrast to the corresponding problem solved by Jackiw and Rebbi for the conventional Dirac theory, where the two components of the fermion field are independent, and, therefore, the zero-energy excitations localized in the mass domain walls are conventional fermion excitations. The momentum space construction of the zero-energy solution presented here is complementary to the approximate real space solution \cite{Sau} of the four-component BdG equations at a vortex in the semiconductor in which superconductivity has been induced by the proximity effect.

\section{Hamiltonian}
The single-particle effective Hamiltonian $H_0$ for the conduction band of a spin-orbit-coupled semiconducting thin film in contact with a magnetic insulator (which induces the Zeeman splitting) is given by ($\hbar=1$) $H_0=H_K + H_{SO} + H_Z$, where,
\begin{eqnarray}
&H_K&=\sum_{\textbf{k},\beta}(\frac{k^2}{2m^*}-\mu)c^{\dagger}_{\textbf{k},\beta}c_{\textbf{k},\beta}\nonumber\\
&H_{SO}&=-\alpha\sum_{\textbf{k},\beta,\gamma}[(\textbf{k}\times\mathbf{\sigma}).\hat{z}]_{\beta,\gamma}
c^{\dagger}_{\textbf{k},\beta}c_{\textbf{k},\gamma} + {\rm{H. c.}}\nonumber\\
&H_Z&=V_z\sum_{\textbf{k},\beta,\gamma}(\sigma_z)_{\beta,\gamma}
c^{\dagger}_{\textbf{k},\beta}c_{\textbf{k},\gamma}
\label{H_0}
\end{eqnarray}
Here, $m^*$, $V_z$ and $\mu$ are the conduction-band effective mass of an electron, effective Zeeman coupling induced
by proximity to a magnetic insulator, and chemical potential, respectively. The Zeeman coupling can also be intrinsic
if the film is made of a magnetic semiconductor (e.g., GaMnAs).
We assume that the Zeeman coupling in the semiconductor is the dominant effect of the proximity to the magnetic insulator and
the direct magnetic field induced by the magnetic insulator is neglected.
\begin{figure}[t]
\hspace*{1.0 in}\includegraphics[width=0.65\linewidth]{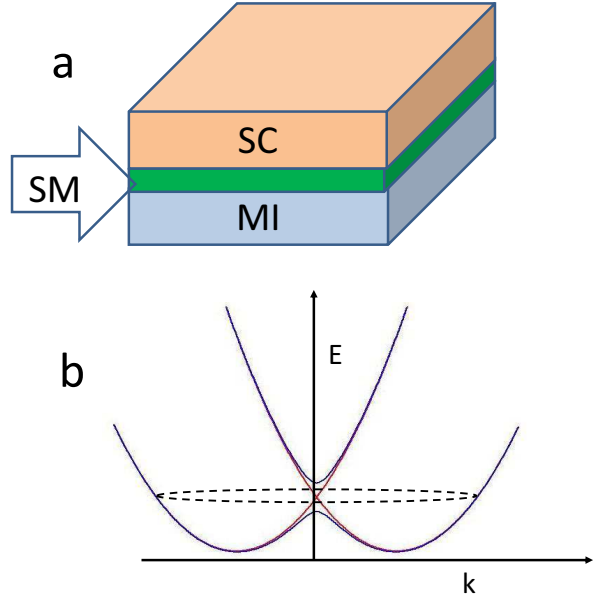}
\caption{(a): The proposed heterostructure of a semiconductor (SM) sandwiched between an $s$-wave superconductor (SC) and a magnetic insulator (MI).
In this geometry, the semiconducting film can support non-Abelian topological order. (b): Single-particle band-structure in the semiconducting film with and without the Zeeman splitting. Without the Zeeman splitting, the two spin-orbit shifted bands touch at $k=0$ (red lines). Then, for any value of the chemical potential, the system has two Fermi surfaces. With a finite Zeeman splitting, the bands have an energy gap near $k=0$ (blue lines).
If the chemical potential lies in the gap, the system just has one Fermi surface (indicated by the dotted circle), a situation conducive to the
emergence of non-Abelian order.}
\label{Fig1}
\end{figure}
 The coefficient
 $\alpha$ describes the strength of the Rashba spin-orbit coupling in the semiconductor, $\beta, \gamma$ are the spin indices, and $\sigma_x, \sigma_y, \sigma_z$ are the Pauli matrices.
The energy spectrum of the spin-orbit-coupled semiconductor
%given by Eq.\eqref{eq:H0}
with $V_z=0$ has two bands crossing the Fermi level.
 This situation should be contrasted with the metallic surface state of a strong TI where an odd number of
 bands cross the Fermi level.
%This is different from the case where an odd number of bands cross the Fermi level in the metallic surface state of the strong
% topological insulator used in previous proposals for realizing Majorana Fermions. It is believed that for time-reversal
%invariant systems the Fermion doubling theorem prevents single bands from crossing the Fermi level.
However,
%by explicitly breaking time-reversal symmetry
by placing the semiconductor in the proximity to a magnetic insulator, it is possible to open a gap in the spectrum, see Fig. (1b). Then,
for $|\mu| < |V_z|$, a single band crosses the Fermi level. This situation is then analogous to the surface of a strong TI in that
an odd number of bands cross the Fermi level ~\cite{fu_prl'08}, and, therefore, is suggestive of supporting non-Abelian
topological order if $s$-wave superconductivity can be induced in the film. This way, by replacing the surface of a 3D strong topological insulator by a regular semiconducting thin-film, the scope of solid state systems which can be designed to support non-Abelian topological order can be vastly expanded \cite{Sau}.

In the presence of an adjacent $s$-wave superconductor,
which induces an $s$-wave pairing potential in the semiconductor by the proximity effect, the full Hamiltonian in the bulk semiconductor becomes,
$H_B=H_0 + H_p$, where,
\begin{equation}
H_p = -\Delta_0 \sum_{\textbf{k}}c^{\dagger}_{\textbf{k},\uparrow}c^{\dagger}_{-\textbf{k},\downarrow} + {\rm{H. c.}}
\label{H_p}
\end{equation}
%In the view of Fu
 %and Kane proposal~\cite{fu_prl'08}, this fact suggests that at the interface of the semiconductor proximity coupled to
%an s-wave superconductor, see Fig.~\ref{fig:structure}, the system might support zero-energy state which can accommodate Majorana fermion.
% This zero-energy Majorana state is stable as long as $|\mu|<|V_z|$
%and disappears when $|\mu|>|V_z|$ indicating that there is a topological phase transition  at $\mu \approx V_z$.
%that if it were possible to introduce a superconducting gap in this system,
%we might have a system that supports a phase with a non-degenerate Majorana state in each vortex for $|\mu|<|V_z|$
%and a regular s-wave superconducing phase without Majorana states for $|\mu|>|V_z|$.
%In the rest of the paper
We show below that the heterostructure in Fig. (1a) has non-Abelian topological order by proving a theorem for the existence of a Majorana zero energy mode at
the core of a vortex with an odd number of flux-quantum in the superconductor.
Our momentum space construction of the zero mode solution in the form of an index theorem is
complementary to the analysis of the real space BdG equations recently carried out \cite{Sau} for the same system.

 In order to describe the spatial dependence of the
superconducting order parameter in the presence of a vortex with winding number $1$, we first write \cite{Sumanta} the
pairing part of the Hamiltonian in real space,
\begin{equation}
H_{V}=-\int d^2R\int d^2r
e^{i\theta_{\bf{R}}}h(R)g({\bf{r}})c^{\dagger}_{{\bf{R}}+{\bf{r}},\uparrow}c^{\dagger}_{{\bf{R}}-{\bf{r}},\downarrow}+
\rm{H.c.}. \label{Pairingvortex}
\end{equation}
Here, $\bf{R}$ and $\bf{r}$ are the center-of-mass and the
relative coordinates of a Cooper pair, respectively. $h(R)$ and
$\theta_{\bf{R}}$ are the amplitude and the phase of the
superconducting order parameter, and, in momentum space, $g(\mathbf{k})$ is the $s$-wave order parameter $\Delta_0$.
In the vortex core,
$h(R)\sim(1-e^{-\frac{R}{\xi}})$ with $\xi$ the coherence length, which describes the suppression of the
condensate amplitude at the core of a vortex.
In order to prove the existence of a non-degenerate zero energy state at the vortex core, we have to map the 2D Hamiltonian $H_T=H_0+H_V$ on the Jackiw-Rebbi problem of 1D Dirac theory with a mass domain wall. Before we do this, let us
briefly discuss the derivation of the zero-mode in the Jackiw-Rebbi problem \cite{Jackiw1} itself.

\section{Fermion zero mode in 1D Dirac theory}
Let us briefly review the derivation of the Jackiw-Rebbi zero
mode for the 1D problem described by the Dirac Hamiltonian,
\begin{equation}
H_{D}=\int
dx\Big[-iv_F \psi^{\dagger} \sigma_z \partial_x \psi
+m(x)\psi^{\dagger}\sigma_x \psi\Big],
\label{h2}
\end{equation}
where $\psi^{\dagger}(x)=(f^{\dagger}_1(x),
f^{\dagger}_2(x))$ with $f_{1,2}(x)$ two
independent fermion fields. Here, $v_F$ is the Fermi velocity of the fermions and $m(x)$ is a spatially-varying mass for the fermion fields.
To uncover the form of the real space wave functions of the quasiparticle excitations, we first write the second-quantized quasiparticle operator,
\begin{equation}
q^{\dagger}=\int
dx~ [\phi_1(x)f^{\dagger}_1(x)+\phi_2(x) f^{\dagger}_2(x)],
\end{equation}
which is assumed to satisfy the equation,
\begin{equation}
[H,q^{\dagger}]=\epsilon q^{\dagger}.
\end{equation}
This implies the
following real-space Dirac equation for the two component wave function
$\phi^{\rm{T}}(x)=(\phi_1(x),\phi_2(x))$:
\begin{equation}
-iv_F\sigma_z\partial_x\phi(x)+\sigma_x m(x)\phi(x)=\epsilon
\phi(x). \label{wa}
\end{equation}
As a first step towards solving this equation, we note that, because $\sigma_y$
anticommutes with $\sigma_x$ and $\sigma_z$, if $\phi(x)$ is an
eigenfunction of Eq.~(\ref{wa}) with eigenvalue $\epsilon$, $\sigma_y\phi(x)$ is
also an eigenfunction of Eq.~(\ref{wa}) with eigenvalue $-\epsilon$. As a result,
if there is a non-degenerate solution of Eq.~(\ref{wa}) with the eigenvalue $\epsilon=0$, it can be made a simultaneous
eigenstate of $\sigma_y$ since $\sigma_y\phi_(x) \propto \phi_(x)$. Let $\phi_0(x)$ denote such a zero-energy solution and
$\sigma_y\phi_0(x)=\lambda\phi_0(x)$. Now, setting $\epsilon=0$ and left-multiplying
Eq.~(\ref{wa}) by $i\sigma_z$ we obtain,
\begin{equation}
\partial_x\phi_0(x)={\lambda\over v_F}
m(x)\phi_0(x),
\end{equation}
which implies
\begin{equation}
\phi_0(x)=e^{{\lambda\over v_F}\int_0^x
m(y)dy}\phi_0(0).\label{zero}
\end{equation}
It can be clearly seen that for $m(x)=\pm {\rm{sign}}(x)|m(x)|$, Eq.~(\ref{zero}) is normalizable for $\lambda=\mp 1$. Therefore, for each sign change of the mass term $m(x)$ in the 1D system, there is a single normalizable zero energy solution. Such a solution indicates the existence of a zero-energy excitation with a quasiparticle operator (for $\lambda = 1$),
\begin{equation}
q^{\dagger}=C\int
dx~ e^{{1\over v_F}\int_0^x
m(y)dy}[f^{\dagger}_1(x)+i f^{\dagger}_2(x)],
\end{equation}
 where $C$ is a normalization constant. Since $f_1(x)$ and $f_2(x)$ are two independent fermion fields, $q^{\dagger}$ above denotes an ordinary fermion operator following fermion anticommutation relations.

Below we map the 2D Hamiltonian $H_T=H_0+H_V$ of the semiconducting thin film with proximity-induced superconductivity on an
effective 1D problem which resembles Eq.~(\ref{h2}), with important differences which
render the quasiparticle mode corresponding to the zero-energy solution a Majorana fermion mode.

\section{Map of the Hamiltonian of the 2D semiconductor on an effective 1D theory}
In order to map the Hamiltonian of the semiconductor, $H_T=H_0 + H_V$, with proximity-induced superconductivity and a vortex at the center, on an effective
one-dimensional problem, we will use the rotational symmetry of $H_T$ to decouple it in the various angular momentum channels \cite{Sumanta}.
We note that, because of the existence of the spin-orbit coupling term $H_{SO}$ in $H_0$,
the Hamiltonian $H_T$ is invariant only under the \emph{simultaneous} rotation
of the system in the real and spin spaces. Therefore, for angular momentum decoupling of $H_T$, we change to a new representation of the fermion operators where they are expanded in the \emph{total} angular momentum channels (as opposed to the orbital angular momentum channels as employed in the case of a chiral \emph{p}-wave superconductor \cite{Sumanta}) indexed by the half-odd-integers $m_J$,
\begin{eqnarray}
&c_{\mathbf{k}\uparrow}&=\frac{1}{\sqrt{2\pi
k}}\sum_{m_J=-\infty}^{\infty}c_{m_J,k,\uparrow}e^{i(m_J-\frac{1}{2})\theta_{\mathbf{k}}}\nonumber\\
&c_{\mathbf{k}\downarrow}&=\frac{1}{\sqrt{2\pi k}}\sum_{m_J=-\infty}^{\infty}c_{m_J,k,\downarrow}e^{i(m_J+\frac{1}{2})\theta_{\mathbf{k}}}
\label{channel},
\end{eqnarray}
where $(m_J \mp 1/2)$ in the exponents on the right indicate the orbital angular momentum quantum numbers for the up and the down spins, respectively. Here, $k=|\mathbf{k}|$ is the magnitude of the vector $\mathbf{k}$, and thus is a one-dimensional variable. The commutation relation $\{c_{\mathbf{k},\beta},c^{\dagger}_{\mathbf{p},\gamma}\}=\delta^2(\mathbf{k}-\mathbf{p})\delta_{\beta,\gamma}$ implies
\begin{equation}
\{c_{m_J,k,\beta},c^{\dagger}_{n_J,p,\gamma}\}=\delta_{m_J,n_J}\delta(k-p)\delta_{\beta,\gamma}.
\end{equation}
Inserting these representations in Eq.~(\ref{H_0}), we find for $H_K$,
\begin{equation}
H_K=\frac{1}{(2\pi)^2}\sum_{m_J}\sum_{\beta}\int dk (\frac{k^2}{2m^*}-\mu)c^{\dagger}_{m_J,k,\beta}c_{m_J,k,\beta}.
\label{HK}
\end{equation}
The Zeeman term, $H_Z$, is also diagonal in the total angular momentum as well as the spin quantum numbers,
\begin{equation}
H_Z=\frac{V_z}{(2\pi)^2}\sum_{m_J}\int dk(c^{\dagger}_{m_J,k,\uparrow}c_{m_J,k,\uparrow}-c^{\dagger}_{m_J,k,\downarrow}c_{m_J,k,\downarrow}).
\label{HZ}
\end{equation}
The spin-orbit term, $H_{SO}$, is diagonal in the total angular momentum quantum number, but not so in the spin quantum number:
\begin{equation}
H_{SO}=\frac{i\alpha}{(2\pi)^2}\sum_{m_J}\int kdk c^{\dagger}_{m_J,k,\uparrow}c_{m_J,k,\downarrow} + {\rm{H.c.}}
\label{HSO}
\end{equation}

Let us now focus on the vortex term $H_V$ in Eq.~(\ref{Pairingvortex}).
Substituting
$$c^{\dagger}_{\bf{R}\pm\bf{r},\beta}=2\pi\sum_{\mathbf{k}}c^{\dagger}_{\mathbf{k},\beta}e^{i\mathbf{k}.(\bf{R}\pm\bf{r})}$$
in Eq.~{\ref{Pairingvortex}}, we end up with two spatial integrals,
$g({\mathbf{k}}-{\mathbf{p}})=\int d^2r
g({\bf{r}})e^{i({\mathbf{k}}-{\mathbf{p}}).{\bf{r}}}=\Delta_0$
%\label{Eq:Integralg
and \begin{equation} I({\mathbf{k}}+{\mathbf{p}})=\int d^2R
e^{i\theta_{\bf{R}}}h(R)e^{i({\mathbf{k}}+{\mathbf{p}}).{\bf{R}}}.\label{I}\end{equation}
%\label{I}\ee
In order to evaluate $I({\mathbf{k}}+{\mathbf{p}})$, we first note that, if in Eq.~(\ref{I}) the vector $(\mathbf{k}+\mathbf{p})$ is rotated
by an angle $\theta$ in the momentum space, the scalar product in the exponent $\exp[i({\mathbf{k}}+{\mathbf{p}}).{\bf{R}}]$ can be made to remain invariant if simultaneously $\theta_{\bf{R}}$  is rotated by $\theta$. Therefore,
\begin{equation}
I(R_{\theta}({\mathbf{k}}+{\mathbf{p}}))=e^{i\theta}I({\mathbf{k}}+{\mathbf{p}}),
\end{equation}
where $R_{\theta}$ is the operator that rotates
$({\mathbf{k}}+{\mathbf{p}})$ by an angle $\theta$ in the momentum space.
It follows that,
$I({\mathbf{k}}+{\mathbf{p}})=e^{i\theta_{{\mathbf{k}}+{\mathbf{p}}}}I(|{\mathbf{k}}+{\mathbf{p}}|)$.
To evaluate $I(|{\mathbf{k}}+{\mathbf{p}}|)$ we choose
$({\mathbf{k}}+{\mathbf{p}})$ along the $y$-axis. Performing the
$\theta_{{\bf{R}}}$ integral which produces $-2\pi i
J_{-1}(|{\mathbf{k}}+{\mathbf{p}}|R)$, where $J_{-1}$ is the Bessel
function of the first kind of order $-1$ \cite{Table}, and then
performing the $R$ integral which produces
$\frac{(2\pi)^3i}{|{\mathbf{k}}+{\mathbf{p}}|^2}\times\mathcal{O}(1)$, we
find,
\begin{equation}
H_{V}=-(2\pi)^3 i\Delta_0\sum_{\mathbf{k},\mathbf{p}
}\frac{ke^{i\theta_{\mathbf{k}}}+pe^{i\theta_{\mathbf{p}}}}{ |{\mathbf{k}}+{\mathbf{p}}|^3 }
c^{\dagger}_{{\mathbf{k}},\uparrow}c^{\dagger}_{{\mathbf{p}},\downarrow}+{\rm{H.c.}}
\end{equation}
Finally, using angular momentum expansion of the fermion operators
and noting that a function of $|{\mathbf{k}}+{\mathbf{p}}|$ is periodic in
$(\theta_{{\mathbf{k}}}-\theta_{{\mathbf{p}}})$ and hence can be Fourier
expanded as
\begin{equation}
\frac{1}{|{\mathbf{k}}+{\mathbf{p}}|^3}=\sum_{m}u_m(k,p)e^{im(\theta_{{\mathbf{k}}}-\theta_{{\mathbf{p}}})}
\end{equation}
with $m$ an integer, we find, after the $\theta_{\mathbf{k}}$ and $\theta_{\mathbf{p}}$ integrals,

\begin{equation}
H_{V}=-i\Delta_0\sum_m\int dkdp \sqrt{kp}u_m(k,p)\Big(k c^{\dagger}_{m+\frac{3}{2},k,\uparrow}c^{\dagger}_{-m-\frac{1}{2},p,\downarrow}
+ p c^{\dagger}_{m+\frac{1}{2},k,\uparrow}c^{\dagger}_{-m+\frac{1}{2},p,\downarrow}\Big).
\label{H_V}
\end{equation}

From Eqs.~(\ref{HK},\ref{HZ},\ref{HSO},\ref{H_V}), it is clear that the $m_J=1/2$ total angular momentum channel separates from the rest (in $H_V$ it is obtained by taking $m=-1$ in the first term and $m=0$ in the second term on the right hand side of Eq.~(\ref{H_V})). It is straightforward to check that no other total angular momentum channel is isolated from the rest. One can also check that for a vortex with an even number of flux quanta, for which $e^{i\theta_{\bf{R}}}$ is replaced by $e^{2i\theta_{\bf{R}}}$ in Eq.~(\ref{Pairingvortex}), no total angular momentum channel can be isolated from the rest of the Hamiltonian. The fact that a single total angular channel can be isolated from the rest is crucial for the following arguments demonstrating the existence of a non-degenerate zero energy solution. The zero energy eigen-solution of the Hamiltonian exists in this channel, and in cases where such an isolated channel does not exist, no non-degenerate zero energy solution exists at the vortex core.

In the $m_J=1/2$ channel, the total Hamiltonian can be written as $H_{T,m_J=\frac{1}{2}}=H_{K,m_J=\frac{1}{2}}+H_{Z,m_J=\frac{1}{2}}+H_{SO,m_J=\frac{1}{2}}+H_{V,m_J=\frac{1}{2}},$ where,
\begin{eqnarray}
&&H_{K,m_J=\frac{1}{2}}=\frac{1}{(2\pi)^2}\sum_{\beta}\int dk (\frac{k^2}{2m^*}-\mu)c^{\dagger}_{\frac{1}{2},k,\beta}c_{\frac{1}{2},k,\beta}\nonumber\\
&&H_{SO,m_J=\frac{1}{2}}=\frac{i\alpha}{(2\pi)^2}\int kdk c^{\dagger}_{\frac{1}{2},k,\uparrow}c_{\frac{1}{2},k,\downarrow} + {\rm{H.c.}}\nonumber\\
&&H_{Z,m_J=\frac{1}{2}}=\frac{V_z}{(2\pi)^2}\int dk(c^{\dagger}_{\frac{1}{2},k,\uparrow}c_{\frac{1}{2},k,\uparrow}-c^{\dagger}_{\frac{1}{2},k,\downarrow}c_{\frac{1}{2},k,\downarrow})\nonumber\\
&&H_{V,m_J=\frac{1}{2}}=-i\int dkdp \Delta(k,p)c^{\dagger}_{\frac{1}{2},k,\uparrow}c^{\dagger}_{\frac{1}{2},p,\downarrow} + {\rm{H.c.}},\nonumber\\
\label{H_HALF}
\end{eqnarray}
where, in the last line, $\Delta(k,p)=\Delta_0\sqrt{kp}(ku_{m=-1}(k,p)+pu_0(k,p))$.
From here onwards we will ignore the subscript $1/2$ in the fermion operators $c, c^{\dagger}$, keeping in mind that all the fermion operators carry the same total angular momentum quantum number $m_J=1/2$.

\section{Demonstration of the zero energy solution}
To turn the Hamiltonian in Eq.~(\ref{H_HALF}) into a form resembling that in Eq.~(\ref{h2}), we have to first diagonalize the free-electron part, $(H_{K,m_J=\frac{1}{2}}+H_{Z,m_J=\frac{1}{2}}+H_{SO,m_J=\frac{1}{2}})$, using a unitary transformation. Only then it is possible to linearize the band energies around the Fermi surfaces and the combined one-electron part in Eq.~(\ref{H_HALF}) may take the Dirac form. To do this, we define the unitary transformation from the $c_{\uparrow/\downarrow,k}^\dagger$ basis to the energy eigenbasis,
\begin{eqnarray}
\left(\begin{array}{c}c_{\uparrow,k}^\dagger\\ c_{\downarrow,k}^\dagger\end{array}\right)&=\left(\begin{array}{cc}a_{k}^*&-b_{k}\\b_{k}^*& a_{k}\end{array}\right)\left(\begin{array}{c}f_{+,k}^\dagger\\ f_{-,k}^\dagger\end{array}\right)\\
\left(\begin{array}{c}f_{+,k}^\dagger\\ f_{-,k}^\dagger\end{array}\right)&=\left(\begin{array}{cc}a_{k}&b_{k}\\-b_{k}^*& a_{k}^*\end{array}\right)\left(\begin{array}{c}c_{\uparrow,k}^\dagger\\ c_{\downarrow,k}^\dagger\end{array}\right)
\end{eqnarray}
where $|a_k|^2+|b_k|^2=1$.
Since the $f_{\pm,k}^\dagger$ operators are related to the original fermion operators by a unitary transformation, they still obey the fermionic anticommutation relations.Calling the corresponding eigenvalues (band energies) $E_{\pm,k}$, the Hamiltonian in this basis in the channel $m_J=\frac{1}{2}$ is written as,
\begin{eqnarray}
H_{T}&=&\int \frac{d k}{(2\pi)^2}[(E_{+,k}-\mu)f_{+,k}^\dagger f_{+,k}+(E_{-,k}-\mu)f_{-,k}^\dagger f_{-,k}]\nonumber\\
&-&\imath \int dk dp \Delta(k,p)[a_{k}^*b_{p}^*f^\dagger_{+,k}f^\dagger_{+,p}-b_{k}a_{p}f^\dagger_{-,k}f^\dagger_{-,p}]\nonumber\\&-&\imath \int dk dp \Delta(k,p)[a_{k}^*a_{p}f^\dagger_{+,k}f^\dagger_{-,p}-b_{k}b_{p}^*f^\dagger_{-,k}f^\dagger_{+,p}] + {\rm{H.c.}}\nonumber\\
&=&\int \frac{d k}{(2\pi)^2}[(E_{+,k}-\mu)f_{+,k}^\dagger f_{+,k}+(E_{-,k}-\mu)f_{-,k}^\dagger f_{-,k}]\nonumber\\
&-&\frac{\imath}{2} \int dk dp [\{\Delta(k,p)a_{k}^*b_{p}^*-\Delta(p,k)a_{p}^*b_{k}^*\}f^\dagger_{+,k}f^\dagger_{+,p}]\nonumber\\&-&\frac{\imath}{2} \int dk dp[\{-\Delta(k,p)b_{k}a_{p}+\Delta(p,k)b_{p}a_{k}\}f^\dagger_{-,k}f^\dagger_{-,p}]\nonumber\\
&-&\imath \int dk dp [\{\Delta(k,p)a_{k}^*a_{p}+\Delta(p,k)b_{p}b_{k}^*\}f^\dagger_{+,k}f^\dagger_{-,p}]
%\nonumber\\&+\frac{\imath}{2} \int dk dp [\{\Delta(k,p)b_{k}b_{p}^*+\Delta(p,k)a_{k}a_{p}^*\}f^\dagger_{-,k}f^\dagger_{+,p}]
+{\rm{H.c.}}
\label{H_T}
\end{eqnarray}
Here, in the second expression, we have written the coefficients of the pairing terms in a form so as to highlight the fact that the coefficients of the intra-band pairing terms must be antisymmetric under the interchange of $k$ and $p$, while the coefficients of the inter-band pairing terms have no such symmetry. The antisymmetry of the coefficients of the intra-band pairing terms, which will be crucial for the demonstration of the zero-energy solution below, follow from the fermion anticommutation relations coupled with the fact that the fermion operators involved are from the same band with the same $m_J$ quantum numbers. Using simplified notations for the coefficients of the pairing terms we rewrite Eq.~(\ref{H_T}) as,
\begin{eqnarray}
H_T&=&\int \frac{d k}{(2\pi)^2}[(E_{+,k}-\mu)f_{+,k}^\dagger f_{+,k}+(E_{-,k}-\mu)f_{-,k}^\dagger f_{-,k}]\nonumber\\
&-&\frac{\imath}{2} \int dk dp [\{\Lambda_{++}(k,p)f^\dagger_{+,k}f^\dagger_{+,p}+\Lambda_{--}(k,p)f^\dagger_{-,k}f^\dagger_{-,p}\nonumber\\
&+&\Lambda_{+-}(k,p)f^\dagger_{+,k}f^\dagger_{-,p}\}] +\rm{H.c.}.
\end{eqnarray}

Let us now assume that the Fermi level is in the lower band $f_{-}$, which has a Fermi momentum $k_F$, and that the upper band $f_{+}$ has a set of
unoccupied states at energy $E_{+,k}-\mu$ separated from the Fermi level by an energy gap $E_+=|V_z|-\mu$ (without loss of generality we take $\mu$ to be positive). This condition is satisfied only as long as $|\mu|<|V_z|$. In what follows we will implicitly assume that this condition is experimentally satisfied. By limiting the momenta in the lower band to those close to the Fermi momentum and in the upper band to those close to the band minimum at $k=0$, the Hamiltonian can be written as,

\begin{eqnarray}
H_T&=&\int \frac{d q}{(2\pi)^2}[E_{+}f_{+,q}^\dagger f_{+,q}+(E_{-,k_F+q}-\mu)f_{-,k_F+q}^\dagger f_{-,k_F+q}]\nonumber\\
&-&\frac{\imath}{2} \int dq dq^{\prime} [\{\Lambda_{++}(q,q^{\prime})f^\dagger_{+,q}f^\dagger_{+,q^{\prime}}+\Lambda_{--}(k_F+q,k_F+q^{\prime})f^\dagger_{-,k_F+q}f^\dagger_{-,k_F+q^{\prime}}
\nonumber\\
&+&\Lambda_{+-}(q,k_F+q^{\prime})f^\dagger_{+,q}f^\dagger_{-,k_F+q^{\prime}}\}] +\rm{H.c}.
\end{eqnarray}

Here, the lower limits on the momenta $q, q^{\prime}$ in the integrals are $0$ or $-k_F$ when the corresponding momentum corresponds to the $+$ or the $-$ bands, respectively (the upper limits on the integrals are taken to be $\infty$).
%Since $q$ and $q^{\prime}$ are assumed to be small compared to $k_F$,
We keep the lowest order dependencies of the functions $\Lambda$ on $q$ and $q^{\prime}$ which are
consistent with the symmetries of these functions from the fermion anticommutation relations, namely that $\Lambda_{++}$ and $\Lambda_{--}$ are antisymmetric in $(q-q^{\prime})$. Since $\Lambda_{+-}(q, q^{\prime})$ has no such symmetry requirement, it is generically non-zero for $q, q^{\prime} \rightarrow 0$, and we assume that it is a non-zero constant $\Lambda_{+-}$ in this limit. This yields,

\begin{eqnarray}
H_T&=&\int \frac{d q}{(2\pi)^2}[E_{+}f_{+,q}^\dagger f_{+,q}+(E_{-,k_F+q}-\mu)f_{-,k_F+q}^\dagger f_{-,k_F+q}]\nonumber\\
&-&\frac{\imath}{2} \int dq dq^{\prime} [\{\Lambda_{++}(q-q^{\prime})f^\dagger_{+,q}f^\dagger_{+,q^{\prime}}+\Lambda_{--}(q-q^{\prime})f^\dagger_{-,k_F+q}f^\dagger_{-,k_F+q^{\prime}}\nonumber\\
&+&\Lambda_{+-}f^\dagger_{+,q}f^\dagger_{-,k_F+q^{\prime}}\}] +\rm{H.c}.
\end{eqnarray}

Let us now define the Fourier transforms to real space,
\begin{eqnarray}
f^\dagger_-(x)&=&\int_{-k_F}^{\infty} dq e^{\imath q x}f^\dagger_{-,k_F+q}\nonumber\\
f^\dagger_+(x)&=&\int_{0}^{\infty} dq e^{\imath q x}f^\dagger_{+,q}.
\label{FT}
\end{eqnarray}
Using Eq.~(\ref{FT}), we fourier transform $H_T$ to real space,

\begin{eqnarray}
H_T&=&\int_{-\infty}^\infty dx dx' [E_{+}\delta(x-x')f_{+}^\dagger(x) f_{+}(x')+E_{-}(x-x')f_{-}^\dagger(x) f_{-}(x')]\nonumber\\
&-&\frac{\imath}{2} \int dx [\{\Lambda_{++}(x)f^\dagger_{+}(x)f^\dagger_{+}(-x)+\Lambda_{--}(x)f^\dagger_{-}(x)f^\dagger_{-}(-x)\nonumber\\
&+&\Lambda_{+-}\delta(x)f^\dagger_{+}(x)f^\dagger_{-}(-x)\}] +\rm{H.c.},
\label{H_Treal}
\end{eqnarray}

where the functions $\Lambda_{++}(x)$ and $\Lambda_{--}(x)$ are odd in $x$, and  $E_{-}(x-x')=\imath v \partial_x \delta(x-x')$ with $v$ the fermi velocity in the lower band.
The Hamiltonian in Eq.~(\ref{H_Treal}) is still not in the same form as in Eq.~(\ref{h2}), so the Jackiw-Rebbi construction does not yet apply. However, as we show below, the BdG equations for a putative zero-energy excitation in the vortex core are of the same form as in Eq.~(\ref{wa}) (with $\epsilon=0$), guaranteeing the existence of a zero energy solution.

Let us define the quasiparticle operator for the putative zero energy mode as,

\begin{equation}
\gamma^\dagger=\int_{-\infty}^\infty dx (\eta_{+,1}(x)f^\dagger_+(x)+\eta_{-,1}(x)f^\dagger_-(x)+\eta_{+,2}(x)f_+(-x)+\eta_{-,2}(x)f_-(-x)).
\label{QP}
\end{equation}

The corresponding BdG differential equations can be derived by setting
\begin{equation}
[H_T, \gamma^{\dagger}]=0.
\label{BdG1}
\end{equation}
To derive the commutators involved in Eq.~(\ref{BdG1}), we need the following anticommutation relations,
\begin{eqnarray}
\{f^\dagger_+(x),f_+(x')\}&=&\int_{0}^\infty dk dk' e^{\imath (kx-k'x')}\{f^\dagger_{+,k},f_{+,k'}\}\nonumber\\
&=&\int_{0}^\infty dk e^{\imath k(x-x')}\equiv S_+(x-x')\nonumber\\
&=&\delta(x-x')/2+\imath P(\frac{1}{x-x'})\\
\{f^\dagger_-(x),f_-(x')\}&=&\int_{-k_F}^\infty dk dk' e^{\imath (kx-k'x')}\{f^\dagger_{-,k},f_{-,k'}\}\nonumber\\
&=&\int_{-k_F}^\infty dk e^{\imath k(x-x')}\equiv S_-(x-x'),
\end{eqnarray}
where the functions $S_+$ and $S_{-}$ are not exactly $\delta$ functions. The existence of $S_{+}$ and $S_{-}$ necessitates the introduction of a new set of four functions,
\begin{equation}
\xi_{\nu,n}(x)=\int_{-\infty}^\infty S_\nu(x_1-x)\eta_{\nu,n}(x_1) d x_1,
\label{new-function}
\end{equation}
where $\nu = \pm$ and $n$ takes the values $1, 2$.

In terms of the $\xi$ functions, the BdG equations take the simple form,
\begin{eqnarray}
\imath v \partial_x \xi_{-,1}(x)-\imath\Lambda_{--}(x)\xi_{-,2}(x)-\imath\Lambda_{+-}\delta(x)\xi_{+,2}(x)/2&=&0\nonumber\\
E_{+}\xi_{+,1}(x)-\imath \Lambda_{++}(x)\xi_{+,2}(x)+\imath \Lambda_{+-}\delta(x)\xi_{-,2}(x)/2&=&0\nonumber\\
-\imath v \partial_x \xi_{-,2}(x)+\imath\Lambda_{--}(x)\xi_{-,1}(x)-\imath\Lambda_{+-}\delta(x)\xi_{+,1}(x)/2&=&0\nonumber\\
-E_{+}\xi_{+,2}(x)+\imath \Lambda_{++}(x)\xi_{+,1}(x)+\imath \Lambda_{+-}\delta(x)\xi_{-,1}(x)/2&=&0
\label{BdG2}
\end{eqnarray}
Eliminating the amplitudes in the upper band, $\xi_{+,1}(x)$, $\xi_{+,2}(x)$, using

\begin{eqnarray}
&\left(\begin{array}{c}\xi_{+,1}(x)\\\xi_{+,2}(x)\end{array}\right)=-\frac{\imath \Lambda_{+-}\delta(x)}{2(E_{+}^2+\Lambda_{++}^2(x))}\left(\begin{array}{cc}E_{+}&-\imath \Lambda_{++}(x)\\\imath\Lambda_{++}(x)&-E_{+}\end{array}\right)\left(\begin{array}{c}\xi_{-,2}(x)\\\xi_{-,1}(x)\end{array}\right)
\label{eliminate}
\end{eqnarray}

we get the equations for the amplitudes in the lower band,

\begin{eqnarray}
\imath v \partial_x \xi_{-,1}(x)&+&\frac{1}{4}
\frac{\Lambda_{+-}^2\delta^2(x) E_{+}}{E_{+}^2+\Lambda_{++}^2(x)}\xi_{-,1}(x)-\imath\left(\Lambda_{--}(x)+\frac{1}{4}
\frac{\Lambda_{+-}^2\delta^2(x)\Lambda_{++}(x)}{E_{+}^2+\Lambda_{++}^2(x)}\right)\xi_{-,2}(x)=0\nonumber\\
-\imath v \partial_x \xi_{-,2}(x)&-&\frac{1}{4}\frac{\Lambda_{+-}^2\delta^2(x) E_{+}}{E_{+}^2+\Lambda_{++}^2(x)}\xi_{-,2}(x)+\imath\left(\Lambda_{--}(x)+\frac{1}{4}
\frac{\Lambda_{+-}^2\delta^2(x)\Lambda_{++}(x)}{E_{+}^2+\Lambda_{++}^2(x)}\right)\xi_{-,1}(x)=0\nonumber\\
\end{eqnarray}

It can now be easily seen that by a redefinition,
\begin{equation}
\psi_{n}(x)=e^{-\imath\frac{1}{4 v}\int_0^x dx' \frac{\Lambda_{+-}^2\delta^2(x') E_{+}}{E_{+}^2+\Lambda_{++}^2(x')}}\xi_{-,n}(x),
\label{Psi}
\end{equation}
the above BdG equations can be recast as,
\begin{eqnarray}
& v \partial_x \psi_{1}(x)+\left(\Lambda_{--}(x)+\frac{1}{4}\frac{\Lambda_{+-}^2\delta^2(x)\Lambda_{++}(x)}{E_{+}^2+\Lambda_{++}^2(x)}\right)\psi_{2}(x)=0\nonumber\\
& v \partial_x \psi_{2}(x)+\left(\Lambda_{--}(x)+\frac{1}{4}\frac{\Lambda_{+-}^2\delta^2(x)\Lambda_{++}(x)}{E_{+}^2+\Lambda_{++}^2(x)}\right)\psi_{1}(x)=0
\label{BdG3}
\end{eqnarray}
By defining the quantity in the parentheses in Eq.~(\ref{BdG3}) as $m(x)$, it can be easily checked that $m(x)$ is an odd function, $m(-x)=-m(x)$, since both $\Lambda_{++}(x)$ and $\Lambda_{--}(x)$ are odd under $x \rightarrow -x$.
Eq.~(\ref{BdG3}) can now be written in terms a two-component spinor wave function $\zeta^{\rm{T}}(x)=(\psi_1(x),\psi_2(x))$,
 \begin{equation}
 v\partial_x\left(\begin{array}{c}\psi_1(x)\\\psi_2(x)\end{array}\right)=-m(x)\sigma_x\left(\begin{array}{c}\psi_1(x)\\\psi_2(x)\end{array}\right).
 \label{BdG4}
 \end{equation}
 which, after multiplication by $\sigma_z$, is of a form similar to Eq.~(\ref{wa}). Therefore, a unique zero energy eigen-solution of $H_T$ in the $m_J=\frac{1}{2}$ channel is guaranteed.

 Notice that, in the above construction for the zero-energy solution of the BdG equations, no special form for the real-space profile of the superconducting gap near the vortex core has been assumed. In particular, we have not assumed the frequently-used step function profile for the gap \cite{tewari_prl'2007, Sau} near the core, without which the real space solution of the second-order, coupled, BdG equations for the present system is still an open problem. All we have needed here to prove the existence of the zero-energy solution, irrespective of the real space profile of the gap, is the antisymmetry of the functions $\Lambda_{++}(x)$ and $\Lambda_{--}(x)$ under $x \rightarrow -x$, which is a consequence of the fermion anticommutation relations. Note that, to use the fermion anticommutation relations to determine the symmetry of these functions, it is crucial that a single total angular momentum channel ($m_J=1/2$) could be isolated from the rest. In the cases where this is not possible, \emph{e.g.}, a vortex with an even number of flux quanta, the Jackiw-Rebbi construction does not apply and a non-degenerate zero-energy solution is not expected at the vortex core.

 It is worth pointing out that the above construction for the zero-energy solution can be straightforwardly applied to describe the zero modes on the surface of a 3D strong TI, described by a Dirac-like Hamiltonian, in the presence of an $s$-wave superconducting vortex \cite{fu_prl'08, Rossi}, but only for a non-zero chemical potential. For zero chemical potential, our reduction of the set of $4$ coupled BdG differential equations to a set of $2$ (with the help of Eq.~(\ref{eliminate})) to bring them to the form in Eq.~(\ref{wa}) in terms of $2 \times 2$ Pauli spin matrices no longer applies. Therefore, our methods above do not directly apply to the problem of zero modes in a TI in the presence of an $s$-wave superconductor at $\mu=0$. However, for this special case, there exists \emph{exact} real space solution of the zero-mode eigenfunction \cite{fu_prl'08, Rossi}.

\section{Demonstration of the Majorana condition}
Now we show that the quasiparticle operator for the zero energy solution found above satisfies the Majorana condition $\gamma^{\dagger}=\gamma$. From Eq.~(\ref{QP}) it is clear that such a condition is satisfied provided
\begin{eqnarray}
&&\eta_{-,1}^{\ast}(-x)=\eta_{-,2}(x)\nonumber\\
&&\eta_{+,1}^{\ast}(-x)=\eta_{+,2}(x)
\label{condition}
\end{eqnarray}
Using Eq.~(\ref{new-function}) and the property for the $S_{\pm}$ functions,
\begin{equation}
S_{\pm}^{\ast}(x_1-x)=S_{\pm}(x-x_1)
\end{equation}
we see that the condition in Eq.~(\ref{condition}) translates into
\begin{eqnarray}
&&\xi_{-,1}^{\ast}(-x)=\xi_{-,2}(x)\nonumber\\
&&\xi_{+,1}^{\ast}(-x)=\xi_{+,2}(x)
\label{condition-xi}
\end{eqnarray}

Now, from Eq.~(\ref{BdG4}) it is clear that $\zeta(x)$ can be taken as an eigen-spinor of $\sigma_x$:
\begin{equation}
\sigma_x \zeta(x)=\lambda \zeta(x)
\end{equation}
Then, solving for Eq.~(\ref{BdG4}), we find that,
\begin{eqnarray}
&&\psi_1(x)=\exp(-\frac{\lambda}{v}\int_0^x m(y)dy)\nonumber\\
&&\psi_2(x)={\rm{sign}}(\lambda)\exp(-\frac{\lambda}{v}\int_0^x m(y)dy),
\end{eqnarray}
where, for $m(x)=\pm {\rm{sign}}(x)|m(x)|$, $\lambda=\pm 1$, in which case the zero mode solution is normalizable. Here we take $\lambda=1$ for the purpose of illustration. Using Eq.~(\ref{Psi}), we get the solutions for the functions $\xi_{-,1}, \xi_{-,2}$,

\begin{equation}
\xi_{-,1}(x)=\exp({\imath\frac{1}{4 v}\int_0^x dx' \frac{\Lambda_{+-}^2\delta^2(x') E_{+}}{E_{+}^2+\Lambda_{++}^2(x')}})\exp(-\frac{\lambda}{v}\int_0^x m(y)dy)=\xi_{-,2}(x)
\label{solution}
\end{equation}

Now, with the help of the phase factor in Eq.~(\ref{solution}) being even in $x$ and
the mass $m(x)$ being odd in $x$, it can be easily checked that
the first line in Eq.~(\ref{condition-xi}) is satisfied. Subsequently, with the help of Eq.~(\ref{eliminate}), and recalling that
$\Lambda_{++}(x)$ is odd in $x$, the second line of Eq.~(\ref{condition-xi}) is also satisfied. Therefore, the quasiparticle operator corresponding to the zero mode solution in Eq.~(\ref{solution}) is a Majorana fermion operator.

\section{Topological protection and topological quantum phase transition}
The proof of the index theorem presented above relies
 on the rotational invariance of the non-interacting part of the Hamiltonian as well as of the vortex
profile, together with the assumption that a single band (lower band, $f_{-}$) dominates the wave-function of the Majorana bound state. The latter assumption is
valid in the limit $E_+=|V_Z|-\mu\gg\Delta_0$. However, as shown below, the question of the existence of the non-degenerate Majorana mode in this system is
 robust to deviations from such assumptions, and the Majorana state can only be removed by a bulk phase transition in the system. Thus,
a non-degenerate Majorana bound state is topologically protected.

In order to see the topological protection of the Majorana mode, let us consider an isotropic BCS Hamiltonian $H_1$
with a symmetric vortex and a small value of $\Delta_0$ (basically, $H_1$ is the same as $H_T$), for which the derivation presented above can be used to show the existence
of a non-degenerate Majorana mode bound to a vortex with an odd number of flux quantum. In addition, let us consider a second BCS Hamiltonian $H_2$ which is possibly
anisotropic (\emph{i.e}., breaks rotational symmetry) near the vortex and has a strong pairing potential.  Therefore, $H_2$ does not lend itself to a simple analytic treatment as given above.  Given
 these two Hamiltonians $H_1$ and $H_2$, we can construct a family of BCS Hamiltonians parameterized by $0<\lambda<1$ given by
$H(\lambda)=\lambda H_1+(1-\lambda)H_2$. Since both $H_1$ and $H_2$ approach the bulk Hamiltonian away from the vortex core, so does $H(\lambda)$.
Therefore, as long as the superconductor has a non-zero bulk gap, $E_g$, away from the vortex core, we expect all the eigenstates of $H(\lambda)$ with energy
less than $E_g$ to be localized at the vortex core. Furthermore, since only a finite number of states can be
localized near the vortex core, there should be a discrete set of eigenstates with energies satisfying $|E_{(s(\lambda)+n)}(\lambda)|<E_g$
 bound to the vortex core. Here, in analogy with a $p$-wave superconductor/superfluid \cite{Kopnin}, $(s(\lambda)+n)$ is an index for the states with $E_{(s(\lambda)+n)}\propto s(\lambda)+n $, where $n$ is an integer and $s(\lambda)=0,1/2$. Note that, if for some value of $\lambda$, $s(\lambda)=0$, a single zero energy state is allowed. Conversely, if $s(\lambda)=1/2$, no zero mode is allowed. In general,
 the shift $s(\lambda)$ can be taken such that it represents an even number (including $0$) zero modes by $s(\lambda)=1/2$
and an odd number of zero modes by $s(\lambda)=0$.
 The quasiparticle excitations of $H(\lambda)$ are given by the equation,
\begin{equation}
[H(\lambda),d_{(s(\lambda)+n)}^\dagger(\lambda)]=E_{(s(\lambda)+n)}(\lambda)d^\dagger_{(s(\lambda)+n)}(\lambda).
\end{equation}
where $d^{\dagger}, d$ are taken as the quasiparticle creation and annihilation operators.
By considering the Hermitian conjugate of the above equation, it can be seen that the solutions of $H(\lambda)$ are paired such that
$d_{-(s(\lambda)+n)}^\dagger(\lambda)=d_{(s(\lambda)+n)}(\lambda)$ and
$E_{-(s(\lambda)+n)}(\lambda)=-E_{(s(\lambda)+n)}(\lambda)$. In the case where $H(\lambda)$ has a single zero energy solution, $(s(\lambda)=0)$, it is clear that  the corresponding zero mode $d_0$ is a Majorana mode $(d_0^\dagger=d_0)$. Since $H_1$ in our problem is assumed to have a non-degenerate Majorana mode (as we have shown by analyzing $H_T$),
$s(\lambda=1)=0$. The stability of this non-degenerate Majorana
zero mode to small perturbations in the BdG Hamiltonian follows from the fact that as long as the low energy states in the vicinity of the vortex core remain discrete in energy spacing, the eigenstates
labeled by  the index $s(\lambda)+n$ evolve continuously with $\lambda$, and thus the shift $s(\lambda)$ cannot jump from it's value $0$ to it's
other allowed value $1/2$. Thus, $s(0)=0$, which implies that $H_2$ has a non-degenerate Majorana
mode as well, even though $H_2$ itself does not lend itself to a simple analysis.

From the above argument it is clear that a non-degenerate Majorana bound state at a vortex remains protected as long as the low energy
states in the vicinity of the vortex core remain discrete in number for all values of the parameter  $0<\lambda<1$. Thus the Majorana bound state
is robust to all local deformations of the Hamiltonian in the vicinity of the vortex core since these cannot affect the energy
gap away from the vortex. On the other hand, if changing the parameter $\lambda$ changes the bulk parameters of the Hamiltonian, such as
$\alpha$, $\mu$ or $V_Z$, then it is possible for the energy gap to close away from the vortex core, causing the states bound to
the vortex core to become delocalized. This results in a breakdown of the above argument for the topological stability of the
Majorana bound states. Thus, the presence or absence of a non-degenerate Majorana bound state at a vortex core leads to a classification
of the phase diagram of two-dimensional superconductors with spin-orbit coupling, such that the phases which support non-degenerate
Majorana bound states are separated from the phases that do not by a quantum phase transition (QPT) where the bulk energy gap closes.

The quantum phases on the two sides of the above QPT can be determined by studying a deformation
parameter $\lambda$ that controls the proximity-induced pairing potential in the semiconductor, \emph{i.e.}, the pairing potential in $H(\lambda)$ is given by $\lambda \Delta_0$.
 The pairing potential increases from $0$ ($\lambda=0$) to a maximum value $\Delta_0$ ($\lambda = 1$). The proof presented above
shows that for $|\mu|<|V_Z|$, for which there is a single band crossing the Fermi-level, and for $\lambda\Delta_0\ll E_+=(|V_Z|-\mu)$, there
is a non-degenerate Majorana mode bound to the core of a vortex. From the bulk Hamiltonian $H(\lambda)$, it is easy to check that the energy gap in the bulk vanishes at a critical value \cite{Sau, Sato},
\begin{equation}
\Delta_{c}=\lambda_c\Delta_0=\sqrt{V_Z^2-\mu^2}.
\end{equation}
Thus our argument for the topological protection of the Majorana bound state implies that a Majorana bound state
 exists at the vortex core in $H(\lambda)$ as long as $\lambda<\lambda_c$ or $\Delta<\Delta_{c}$. Moreover, this phase
is separated from the phase with $\Delta>\Delta_c$ by a QPT at which the single particle energy gap vanishes. The spin-dependent terms in the Hamiltonian, such as the Zeeman splitting and the spin-orbit coupling, cease to play a critical role in the phase $\Delta>\Delta_c$, and it is possible
to reduce both these couplings to zero without crossing another phase transition (\emph{i.e.}, without $E_g$ going through $0$ again). Thus, the phase $\Delta>\Delta_c$ must be in the same class as
a conventional $s$-wave superconductor without spin-orbit coupling or Zeeman splitting, and therefore cannot support a non-degenerate Majorana mode.
Furthermore, the symmetries of the superconducting order parameters are identical in these two phases.
%do not show any singular changes as the pairing potential $\Delta$ crosses the critical value
%$\Delta_c$.
%Therefore it can be said that phase diagram of the 2 dimensional spin-orbit coupled semiconductor is composed of 2 topological
%phases that are characterized by the presence or absence of Majorana bound states at the cores of vortices and
Therefore, these two phases, distinguished by the presence or absence of a Majorana mode, are separated
by a topological QPT at $\Delta=\sqrt{V_z^2-\mu^2}$ where the energy gap closes.

\section{Conclusion}
We prove a theorem for the existence of Majorana zero modes in a semiconducting thin film with a sizable spin-orbit coupling, in which $s$-wave superconductivity and a Zeeman splitting are induced by proximity effect (Fig. (1)). Our momentum-space construction of the zero-mode solution in the form of the Jackiw-Rebbi index theorem \cite{Jackiw1}, which is complementary to the approximate real-space solution of the BdG equations at a vortex core \cite{Sau}, proves the existence of non-degenerate Majorana fermion excitations localized at the vortices in the semiconductor heterostructure shown in Fig. (1a).

In our method, no special form (in particular, the frequently-used step-function form \cite{tewari_prl'2007,Sau}) for the real-space profile of the superconducting pairing potential near the vortex core is required. We use only the antisymmetry property of the intra-band pairing potentials, $\Lambda_{++}(x)$ and $\Lambda_{--}(x)$ (Eq.~(\ref{H_Treal})) under $x \rightarrow -x$, which is a consequence of the fermion anticommutation relations. Therefore, the non-degenerate Majorana mode found here is robust to local spatial deformations of the order parameter. For the construction in this paper to apply, it is crucial that a single total angular momentum channel (for a vortex with a single flux quantum: $m_J=1/2$) can be isolated from the rest of the Hamiltonian. In the cases where this is not possible, \emph{e.g.}, for a vortex with an even number of flux quanta, the theorem does not apply and a non-degenerate zero-energy solution is not expected at the vortex core. The methods of this paper can be straightforwardly applied to the zero modes on the surface of a TI with a non-zero chemical potential in the presence of a superconducting vortex \cite{fu_prl'08}, but not to the case when the chemical potential on the TI surface is zero. However, in the latter special case of zero chemical potential, an exact solution of the BdG equations already exists \cite{fu_prl'08, Rossi} for the zero-mode eigenfunction for arbitrary real-space profile of the order parameter.

\section*{Acknowledgements}

This work is supported by DARPA-QuEST, JQI-NSF-PFC, and LPS-NSA.
We thank D. H. Lee for previous collaboration on a related work in chiral $p$-wave superconductors \cite{Sumanta}.
ST acknowledges DOE/EPSCoR Grant \# DE-FG02-04ER-46139 and Clemson University start up funds for support.

%\section*{References}

%\thebibliography{99}

\end{document}